# Quantitative test of the barrier nucleosome model for statistical positioning of nucleosomes up- and downstream of transcription start sites

Wolfram Möbius and Ulrich Gerland*

Arnold Sommerfeld Center for Theoretical Physics (ASC) and Center for NanoScience (CeNS), LMU München, Theresienstraße 37, 80333 München, Germany

∗ E-mail: gerland@lmu.de

## Abstract

The positions of nucleosomes in eukaryotic genomes determine which parts of the DNA sequence are readily accessible for regulatory proteins and which are not. Genome-wide maps of nucleosome positions have revealed a salient pattern around transcription start sites, involving a nucleosome-free region (NFR) flanked by a pronounced periodic pattern in the average nucleosome density. While the periodic pattern clearly reflects well-positioned nucleosomes, the positioning mechanism is less clear. A recent experimental study by Mavrich *et al.* argued that the pattern observed in *Saccharomyces cerevisiae* is qualitatively consistent with a 'barrier nucleosome model', in which the oscillatory pattern is created by the statistical positioning mechanism of Kornberg and Stryer. On the other hand, there is clear evidence for intrinsic sequence preferences of nucleosomes, and it is unclear to what extent these sequence preferences affect the observed pattern. To test the barrier nucleosome model, we quantitatively analyze yeast nucleosome positioning data both up- and downstream from NFRs. Our analysis is based on the Tonks model of statistical physics which quantifies the interplay between the excluded-volume interaction of nucleosomes and their positional entropy. We find that although the typical patterns on the two sides of the NFR are different, they are both quantitatively described by the same physical model, with the same parameters, but different boundary conditions. The inferred boundary conditions suggest that the first nucleosome downstream from the NFR (the +1 nucleosome) is typically directly positioned while the first nucleosome upstream is statistically positioned via a nucleosome-repelling DNA region. These boundary conditions, which can be locally encoded into the genome sequence, significantly shape the statistical distribution of nucleosomes over a range of up to $\sim$ 1000 bp to each side.

## Author Summary

Within the last five years, knowledge about nucleosome organization on the genome has grown dramatically. To a large extent this has been achieved by an increasing number of experimental studies determining nucleosome positions at high resolution over entire genomes. Particular attention has been paid to promoter regions, where a canonical pattern has been established: a nucleosome free region with pronounced adjacent oscillations in the nucleosome density. Here we tested to what extent this pattern may be quantitatively described by a minimal physical model, a one-dimensional gas of impenetrable particles, commonly referred to as the 'Tonks gas'. In this model, density oscillations occur close to a boundary at dense packing. Our systematic quantitative analysis reveals that, in an average over many promoters, a Tonks gas model can indeed account for the nucleosome organization to both sides of the nucleosome free region, if one allows for different boundary conditions at the two edges: On the downstream side, a single nucleosome is typically directly positioned such that it forms an obstacle for the neighboring nucleosomes, while such a barrier nucleosome is typically missing on the upstream side.

# Introduction

The long DNA molecules of eukaryotic genomes are packaged into a compact structure with the help of histone proteins [1]. The fundamental unit of this structure, a nucleosome, comprises almost 150 base pairs (bp) of DNA wrapped around a histone octamer [2,3]. Individual nucleosomes are typically linked by 15 - 70 bp of free DNA into a "beads on a string" conformation, the primary and most stable structural level of chromatin. While packaging renders the genome compact, it also makes up to 80 % of the DNA inaccessible for protein-binding at any given time [4], potentially hindering the molecular processing of genetic information. In principle, accessibility might be attained dynamically, since mechanisms are known for spontaneous unwrapping [5,6] and diffusive sliding of nucleosomes [7], as well as active remodeling [8]. However, numerous recent studies indicate that nature's solution to the accessibility issue is based, at least in part, on the widespread use of nucleosome positioning [4,9–14]. Nucleosome positioning essentially amounts to the opposite strategy of constraining the mobility of nucleosomes, rendering a selected set of DNA sites constantly accessible.

Recent experiments measuring the distribution of nucleosomes across the genomes of several model organisms have robustly identified three salient features [11]: (i) A significant fraction of nucleosomes appears rather well positioned. In other words, the nucleosome positions determined from a large ensemble of cells do not average out to a constant density, but display many pronounced peaks. (ii) Typically, genes have a nucleosome-free region (NFR) upstream of their transcription start site (TSS). That is, when genes are aligned at the TSS and with the direction of transcription to the right, the average nucleosome density exhibits a clear dip, about one nucleosome wide, to the left of the TSS. (iii) Downstream of the TSS, the gene-averaged nucleosome density displays strong oscillations, with an amplitude that decays with the distance from the TSS. Furthermore, biochemical experiments have firmly established that the DNA-binding affinity of histones depends on the DNA sequence, largely due to the intrinsic sequence-dependence in the biophysical properties of DNA, such as its bendedness and bendability [15]. Hence, a genomic free energy landscape for nucleosome positioning can be programmed into the genome sequence by appropriate placement of nucleosome attracting and repelling sequence motifs.

Indeed, bioinformatic and biophysical approaches that parameterize sequence-encoded effects on nucleosome positioning have been remarkably successful in modeling and predicting the large-scale genomic nucleosome occupancy [16–20], which has led to the notion of a genomic code for nucleosome positions [16]. Yet, the causes of the three above salient features are not yet disentangled. In particular, a recent study on nucleosome positioning in *Saccharomyces cerevisiae* [10] argued that the oscillatory pattern in the average nucleosome organization downstream of the TSS is qualitatively consistent with the statistical positioning mechanism proposed by Kornberg and Stryer [21]. With this mechanism, most nucleosomes are not individually positioned, but a non-random relative arrangement arises collectively, from statistical correlations induced by the interaction between neighboring nucleosomes. The phase of such a statistical arrangement relative to the DNA is determined by "barriers" on the genome, i.e., local disturbances of the "nucleosome gas". A disturbance is created regardless of whether the local effect on nucleosomes is attracting or repelling, e.g., by sequences that attract or repel nucleosomes or by other bound proteins [22,23]. According to this scenario, termed the 'barrier nucleosome model' [10,24], sequence-encoded positioning is required only for barrier creation, whereas nucleosomes adjacent to the barriers are positioned "for free", i.e., primarily via statistical correlations and with DNA sequence playing only a minor role.

However, while the observed oscillatory pattern downstream of the TSS is reminiscent of the pattern calculated by Kornberg and Stryer [21], there should be a similar pattern upstream of the TSS if statistical positioning is indeed the dominant force, since barriers act to both sides. Also, can the observed pattern be quantitatively explained by statistical positioning? Finally, does the precise shape of the pattern permit conclusions on the nature of the barrier, e.g., whether it is caused by an attractive or repelling effect on nucleosomes? Here, we address these quantitative questions using the yeast data of Mavrich *et al.* [10] and a quantitative description of statistical positioning, which is essentially the same as in the work of Kornberg and Stryer [21] and equivalent to the (much older) "Tonks gas" model from statistical physics [17, 25–28].

## Results

### Quantitative barrier nucleosome model

Kornberg noted early on [24] that a nonrandom quasi-periodic nucleosome pattern arises already from the interplay of two basic biophysical constraints, (i) the constraint that the same DNA segment cannot simultaneously be incorporated into two nucleosomes, and (ii) the constraint that nucleosomes cannot form at 'barrier' genome locations, e.g., those already occupied by other proteins such as sequence-specific transcription factors. The significance of the first constraint is that the exclusion between nucleosomes creates correlations in their statistical distribution along the DNA. Theoretically [26], these correlations are revealed by a decaying oscillatory pattern in the two-particle distribution function, $\rho_2(r)$, which measures the probability to find, in the ensemble of all admissible nucleosome configurations, a nucleosome at a given location and another one at a distance $r$ from it. In other words, the knowledge of the position of one nucleosome leads to a partial knowledge of the nucleosome positions in the vicinity (however, this two-particle distribution function is difficult to measure directly in experiments). The significance of the second constraint is that barriers in the "nucleosome gas" pin down the phase of the correlations, such that even the average nucleosome density $\rho(r)$ displays a decaying oscillation as a function of the distance $r$ from the barrier [21]. Such barriers can be created by a variety of mechanisms; in particular, barriers can also be directly encoded in the DNA sequence, e.g., via poly(dA:dT)-tracts that are energetically unfavorable to incorporate into the nucleosome structure [29]. Similarly, "road block" nucleosomes that are particularly well-positioned will form a barrier for the surrounding nucleosomes.

Here, we treat the average nucleosome density $\rho(r)$ as a quantitative experimental feature that can be assayed for clues about the nature of these barriers and, more generally, about the extent to which statistical positioning is reflected in the nucleosome organization *in vivo*. This analysis must be based on a quantitative description of statistical positioning. In statistical physics, the interplay between interaction and entropy of particles in a one-dimensional configuration space has long been quantified in simple models for gas/liquid systems [25, 26, 30]. The classic quantitative study of statistical positioning, by Kornberg and Stryer [21], is also consistent with this general framework. The simplest model is the 'Tonks gas' [25] where particles with a fixed size $b$ and a mean density $\overline{\rho}$ interact only via hard-core repulsion that makes them impenetrable. For this model, the explicit analytical expression for the average particle density at a distance $r$ (in bp) from a perfect barrier is [26]

$$\rho(r) = \sum_{k=1}^{\infty} \frac{(\frac{r}{b}-k)^{k-1}\Theta(\frac{r}{b}-k)}{b\cdot(k-1)!} \left(\frac{\overline{\rho}\,b}{1-\overline{\rho}\,b}\right)^k e^{-\frac{r/b-k}{1/\overline{\rho}b-1}}, \qquad (1)$$

where $\Theta(r)$ denotes the Heaviside step function. This average density is related to the above-mentioned two-particle distribution function in an infinite system via $\rho(r)\cdot\overline{\rho} = \rho_2(r)$. Eq. (1) produces the decaying oscillatory pattern that is characteristic for statistical positioning, see Fig. S1 for an illustration and 'Materials and Methods' for a self-contained derivation and a brief discussion of the physical mechanism underlying the density oscillation. The wavelength of the oscillatory pattern and the characteristic length over which its amplitude decays are both determined by the two physical parameters of the model, i.e., the particle size $b$ and the average particle density $\overline{\rho}$. Note that the expression (1) holds only for a perfect barrier; more general situations will be considered below.

As Eq. (1) describes a nontrivial effect that arises only from properties which the "nucleosome gas" shares with any other one-dimensional gas of impenetrable particles, it can be regarded as a 'null model', i.e., a quantitative reference that helps to identify relevant effects beyond the universal features for systems of this class. With this goal in mind, we wanted to compare Eq. (1) to patterns extracted from experiments.

## Extraction of experimental nucleosome patterns

To extract the consensus distribution of nucleosomes around the NFR at the 5' end of genes, previous studies aligned the genes at their TSS and averaged the nucleosome distributions over all genes [11]. This procedure is not suitable for our quantitative analysis, since the TSS cannot be mapped to a feature in the nucleosome gas. Instead, we used the positions of the NFR-flanking nucleosomes as reference points for our alignments, which permits a quantitative comparison of the averaged pattern with the nucleosome gas model (see below).

In addition to the appropriate choice of reference point for the alignment, it was important to process the experimental data in a way such that it became directly comparable to the physical density $\rho(r)$. Many studies determine nucleosome positions using a procedure of the following type [11]: First, the nucleosomal DNA is extracted from an ensemble of cells using micrococcal nuclease (MNase). The genomic positions of these DNA fragments are then located using hybridization or sequencing approaches. Usually this raw data is further processed with hidden Markov models (e.g., [9]) or peak detection algorithms (e.g., [31]), in order to infer the typical or putative nucleosome positions. These typical nucleosome positions are then used for subsequent analysis of nucleosome organization, including the consensus distribution around NFRs. However, such averages over typical nucleosome positions do not correspond to a physical observable. For qualitative analysis, the data processing algorithms are useful filters to enhance and highlight positioning effects. However, the use of a single, typical position for a nucleosome eliminates any cell-to-cell variation in the position. For our quantitative analysis, we had to use the undistorted raw data instead (i.e., the density of DNA reads along the genome for the sequencing approach), which is the best available experimental proxy to the physical density $\rho(r)$, see 'Materials and Methods' for details. Note that our observable, the nucleosome density, is distinct from the other frequently used observable, the nucleosome occupancy, which measures the probability to find a specified base pair covered by a nucleosome.

Fig. 1 summarizes the nature of the data from a physics perspective. As illustrated in Fig. 1A, the extracted nucleosomal DNA originates from many cells with nucleosome positions that generally differ from cell to cell. The experimentally observed read density corresponds to the histogram shown in the bottom of Fig. 1A. This histogram would be directly comparable to the theoretical density $\rho(r)$ for a nucleosome gas, if (i) the average over the different cells is equivalent to the thermal average, (ii) a DNA read identifies a nucleosome position uniquely and precisely, and (iii) the average number of reads per nucleosome is known and its fluctuations due to the random sampling are negligible. None of these conditions is entirely satisfied. Clearly, the relevant question (discussed in 'Materials and Methods') is how much this affects the physical interpretation of the data. Since the average number of reads per nucleosome is in fact unknown, it is already clear that one cannot readily convert the read density to an absolute nucleosome density, i.e., the experimental proxy to $\rho(r)$ is not normalized. Fig. 1B illustrates the second averaging procedure, which is akin to a "disorder average" in statistical physics, in that it involves averaging over an ensemble of different systems rather than an ensemble of different states of the same system. Clearly, each gene is intrinsically different and could display a distinct pattern of nucleosome organization. However, as illustrated in the bottom of Fig. 1B, the common pattern that emerges by aligning the genes by the position of their +1 nucleosome (the first downstream from the NFR) exposes the generic features in a large set of genes. For individual genes, this pattern is obscured by the noise due to the limited statistics of the data.

We performed our analysis on the data of Mavrich *et al.* [10]. The red dots in Fig. 2 display the average read density when the genes are aligned to the +1 nucleosome, with the direction of transcription from left to right. Our definition of the +1 nucleosome position is the most likely position of the first nucleosome downstream from the TSS based on the list of TSSs and nucleosomes by Mavrich *et al.* [10]; see 'Materials and Methods' for details. On a qualitative level, the pattern of Fig. 2 (red dots) closely resembles the consensus pattern from previous studies (see, e.g., Fig. 2 in Ref. [11]). In particular, both display the same salient features, i.e., the pronounced downstream oscillations, the slow decay to a constant density, the nucleosome-free region, and the weak upstream oscillations. However, on a quantitative level, the patterns are significantly different, and only the pattern of Fig. 2 is suitable for quantitative comparison with a physical model.

Our analysis leading to Fig. 2 did not include a correction for the known sequence bias of the MNase enzyme [32, 33]. However, Fig. S2 compares the pattern of Fig. 2 with the result of an alternative analysis that also incorporates a correction for the MNase bias, and suggests that the MNase bias does not significantly affect the pattern; see 'Materials and Methods' for details. Another concern is that the entire set of genes contains a significant fraction where the gene ends within the 2000 bp downstream range plotted in Fig. 2, see Fig. S3A. Therefore, we repeated our analysis on the subset of long genes with a size of more than 2000 bp in length. Fig. S3C shows that the resulting pattern is quantitatively very similar to that of Fig. 2. Taken together, these results indicate that the pattern of Fig. 2 (red dots) represents a robust quantitative signature of the nucleosome organization near transcription start sites in yeast.

## Quantitative analysis

To interpret the extracted pattern within the physical model described above, we performed a nonlinear least-squares fit to Eq. (1), as described in 'Materials and Methods'. We kept the width of the nucleosomes fixed at the value $b = 147$ bp suggested by the crystal structure [3], and hence the only fit parameters were the mean nucleosome density $\bar{\rho}$ and the global normalization factor for the data (see above). The best fit is displayed as a gray line in Fig. 2A. To judge the quality of the agreement, it is useful to recall that the experimental pattern is basically described by five quantitative characteristics: the period of the oscillation, the length scale over which the oscillation decays, the asymptotic value of the density, and the amplitudes of the peaks and valleys in the density, above and below the asymptotic line. Given only two fitting parameters, the overall quantitative agreement between the physical model and the biological data is therefore remarkably good.

Fig. S4 shows the corresponding fit to only the set of long genes, with a similar result. In both cases, the most apparent deviation between the model and the data is in the shapes and the amplitudes of the first two peaks, associated with the +1 and +2 nucleosome. We wanted to test whether this is solely a consequence of the fact that Eq. (1) assumes a perfectly positioned +1 nucleosome, while the data displays a small residual positional variability for the +1 nucleosome. We therefore convoluted the model density, Eq. (1), with the shape of the +1 peak in the data (see 'Materials and Methods' for details). The corresponding fit of this convoluted density to the data is shown in Fig. 2B (gray line). By construction, the shape of the +1 peak now matches, but we note that the deviation in the +2 peak disappeared as well, suggesting that the finite positional variability of the +1 nucleosome is indeed sufficient to explain most of the deviation between the physical model and the biological data.

Before discussing the obtained parameter values and the robustness of the fitting procedure, we address the immediate question that emerges from the above results: On the one hand, the agreement between model and data is consistent with the hypothesis that most of the nucleosomes downstream of the +1 nucleosome are statistically positioned. On the other hand, the statistical positioning mechanism has no intrinsic bias to a particular direction, i.e., the pattern upstream of the NFR should be described as well by a viable physical model. However, the upstream consensus pattern reported in previous studies displays much less pronounced oscillations than on the downstream side [4, 10]. To test whether this is simply a consequence of the gene-to-gene variation in the distance between the -1 nucleosome and the TSS, which should smear out the averaged pattern, we analyzed the statistical distribution of these distances and realigned all genes by the position of their -1 nucleosome. The -1 position is defined here by the first nucleosome upstream from the TSS, see 'Materials and Methods'.

Fig. 3A displays the statistics of the ±1 nucleosome positions relative to the TSS, as derived from the nucleosome map of Ref. [10]. While +1 nucleosomes are restricted to a region about 50 bp downstream from the TSS [31,34], the -1 nucleosome position is considerably more disperse. Accordingly, the distance between the +1 and -1 nucleosomes, i.e., the gap size, also has a wide distribution, see Fig. 3B. This distribution indeed smears out an oscillatory upstream pattern, which is uncovered by an alignment to the -1 nucleosome position that eliminates the gap size variation, see Fig. 4A (blue dots). However, while this upstream pattern does display regular oscillations, the comparison to the superimposed downstream pattern from Fig. 2 demonstrates that these two patterns are significantly different. A possible concern with this upstream pattern is the frequent occurrence of another NFR closely upstream of the -1 nucleosome, either at the start of a divergently transcribed neighboring gene or at the 3' end of a gene transcribed in the same direction (3' NFRs are analyzed further below). To address this concern, we selected only the subset of genes with no gene start or end within 1000 bp upstream of the TSS and compared their averaged pattern to that for all genes. Fig. S3D shows that these two patterns are quantitatively very similar (and clearly different from the downstream pattern), suggesting that the adjacent NFRs located at various distances have no significant effect on the average upstream pattern.

The difference in the up- and downstream pattern might be an indication of positioning mechanisms beyond statistical positioning. Alternatively, this difference might be due to an intrinsic asymmetry of the NFRs, caused by different molecular determinants for the up- and downstream NFR boundary. Such an asymmetry would lead to a different boundary condition for the nucleosome gas on the two sides of the NFR. To illustrate the possible effect of the boundary condition on the pattern in the nucleosome gas, Fig. 4B shows the patterns for a range of boundary conditions together in a 3D plot. Here, the different boundary conditions are parameterized by an energy scale, $\epsilon_0$, which measures the strength and the sign of the local effective free energy for nucleosome binding: Positive $\epsilon_0$ (towards the front) correspond to a nucleosome repellent region, i.e., nucleosomes at positions to the left of the origin receive an energetic penalty $\epsilon_0$. In contrast, negative $\epsilon_0$ (towards the back) correspond to an attractive positioning potential that is localized to a narrow region, the width of which is chosen here to roughly correspond to the finite peak width of the +1 nucleosome in the data. Note that all of the patterns contained in the 3D plot of Fig. 4B are qualitatively similar, irrespective of the value of $\epsilon_0$. However, they are different on a quantitative level, and we next exploit this difference, using the experimental pattern as a quantitative signature, to infer the type of the boundary condition that is effectively implemented *in vivo*.

In particular, it is instructive to contrast the case of a perfectly repulsive barrier ($\epsilon_0 \to \infty$) with a perfect attractive positioning potential ($\epsilon_0 \to -\infty$). Our above analysis of the downstream pattern in Fig. 2 was based on the latter case, i.e., we assumed that most +1 nucleosomes are directly kept at particular positions on the genome through the action of specific molecular forces. We found that this assumption is compatible with the data. Given that the upstream pattern does not comply with this direct positioning scenario, we hypothesized that most -1 nucleosomes are instead indirectly (statistically) positioned by a repulsive barrier located at the upstream edge of the NFR. Fig. 4C displays the upstream pattern (blue dots) together with the model prediction assuming a perfectly repulsive barrier (gray line). Note that this prediction is obtained with the same values for $\overline{\rho}$ and normalization factor as inferred from the fit to the downstream pattern, i.e., it does not involve parameter fitting, see 'Materials and Methods' for details. The agreement is surprisingly good, consistent with the interpretation that the positioning of most nucleosomes in the vicinity of the TSS is induced by a NFR that is intrinsically asymmetric: Our quantitative comparison suggests that the upstream boundary of the NFR is typically determined by repulsion rather than direct positioning of the -1 nucleosome.

To put these observations on a systematic basis, we performed simultaneous fits on both sides of the TSS, for all combinations of boundary conditions and compared the results quantitatively on the basis of the mean square deviation per data point, see Fig. S5, Table S3, and 'Materials and Methods'. The results corroborate that the experimental pattern is best explained by the scenario where the +1 nucleosome is directly positioned whereas the -1 nucleosome is statistically positioned by a repellent region, as illustrated in Fig. 4D. The second best fit is obtained by the scenario where both the -1 and the +1 nucleosome are statistically positioned.

As Fig. S5 shows, both patterns are quantitatively well explained with a single average nucleosome density $\bar{\rho} = 1/180$ bp for both up- and downstream of the TSS. Indeed, we find no clear evidence in the data that the average density of nucleosomes is different in intergenic and genic regions (see 'Materials and Methods'), contrary to some observations made in other studies. We robustly obtained density values $\bar{\rho}$ in the range of one nucleosome per 172 to 180 bp, described above and independent of the detailed choice of the fitting method. These values are slightly (but consistently) larger than the "nucleosome mode" of 165 bp inferred by Mavrich *et al.* [10] by determining the typical peak to peak distance in the experimental pattern.

Finally, it is interesting to note that NFRs have also been reported at the 3' end of genes, although their biological significance is obscure [10, 35]. In order to see to what extent our findings can be generalized to this class of NFRs, we also extracted the average up- and downstream pattern for 3' NFRs by aligning to the respective flanking nucleosomes. Fig. S6 shows these patterns; see caption for details. We observe that on neither side the pattern displays the strong features associated with the direct positioning scenario. Instead, both 3' patterns resemble the 5' upstream pattern, which is superimposed for comparison in Fig. S6. This suggests that the 3' NFR is typically only a repulsive region, and hence less structured than the typical 5' NFR.

## Discussion

The recent genome-scale identification of nucleosome positions revealed that a large fraction of nucleosomes are non-randomly positioned, that a large fraction of genes have a nucleosome-free region (NFR) at their promoters, and that the NFRs are flanked by salient oscillatory patterns in the nucleosome density [9, 10]. Here, we performed a quantitative analysis of the average up- and downstream patterns, to reveal hidden information about factors that affect nucleosome positioning in promoter regions. To this end, we reanalyzed previously published yeast data [10] in a physical way. We found that the up- and downstream patterns differ significantly, but both are quantitatively consistent with a minimal model where nucleosome positioning is effected only from the location of the NFRs, but radiates over a range of up to $\sim 1000$ bp to each side via the statistical positioning mechanism. Within this model, the difference in the average up- and downstream patterns is explained as an intrinsic asymmetry of the NFRs, which leads to different boundary conditions for the "nucleosome gas" on the two sides, see Fig. 4. In contrast, we found no evidence of such an asymmetry for 3' NFRs at the end of genes.

That statistical positioning in the vicinity of barriers is a mechanism capable of producing a non-random nucleosome arrangement has long been established theoretically [21] and experimentally [22]. Statistical positioning of nucleosomes around promoter regions has been proposed several years ago [9], while testing of this hypothesis has started only very recently [10, 36–38]. The first study [10] presented qualitative evidence for statistical positioning, but was limited by its approach relying on consensus nucleosome positions and TSS alignments. However, that study also performed a thorough statistical analysis of the DNA sequence around promoters, and found that sequence elements known to be involved in nucleosome positioning (dinucleotide patterns and poly(dA:dT) stretches) are concentrated to the NFR and the positions of the -1 and +1 nucleosomes, and are significantly less frequent up- and downstream from this region. This finding is consistent with our conclusions drawn from the quantitative analysis of the nucleosome patterns. Additionally, our analysis suggests that the sequence elements around the position of the -1 nucleosome are either not sufficiently widespread or not sufficiently effective to directly position the -1 nucleosome in the average pattern. This is not unlikely given that other mechanisms than direct sequence specificity are needed to obtain the precise positioning of the +1 nucleosome *in vivo* [39].

Two additional studies on statistical positioning in genic regions appeared after our work was completed [36, 37]. These studies did not consider alignments to TSSs or +1 nucleosomes, but instead ranked genes by the distance between their first and last nucleosome, revealing a striking organization of the local minima in the nucleosome occupancy. This organization was found to be consistent with a Tonks gas that is constrained by repelling barriers from both sides. This analysis, with its focus on the genic regions and the positions of the minima, is complementary to ours, which focused on the quantitative shape of the average density, in particular also in the upstream intergenic region, and analyzed the difference between the up- and downstream pattern.

Taken together, our and the existing studies of statistical positioning support the view that long-range correlations in nucleosome positions produced by localized features in the effective free energy landscape for nucleosome binding are an important determinant of the genome-wide nucleosome organization. Indeed, for yeast, where TSSs are typically spaced $< 2000$ bp apart (Fig. S3B), statistical positioning from features encoded only at the TSSs is sufficient to obtain non-random positioning for most nucleosomes. The physical origin of statistical positioning is an interplay between the mutual exclusion and the positional entropy of nucleosomes. While this mechanism does not "glue" nucleosomes to specific locations on the DNA, it does effect that, on average, nucleosomes favor certain positions over others. It can therefore make specific (binding) sites on the DNA more (or less) accessible for proteins. Moreover, it may also cause a bias for mutation processes, thereby creating a position-dependent mutation rate [40] and possibly long-range DNA sequence correlations.

The approach taken in the present study may be classified as a "reverse approach", which starts from the observed distribution of nucleosomes along the genome and ultimately seeks to determine from it the underlying free energy landscape for nucleosome binding (see 'Materials and Methods' for a discussion of the assumptions leading to the concept of an effective free energy landscape). Here, this approach has led to the typical form of local features in the landscape that is depicted in Fig. 4D. Note that by construction, our approach has two important limitations:

First, it cannot pinpoint the molecular mechanisms responsible for creating the features in the effective free energy landscape. For instance, our findings are equally compatible with sequence-determined depletion like in the *HIS3-PET56* promoter [23], chromatin remodeler induced nucleosome organization like in the *POT1* promoter in its repressed state [41], or with varying promoter architecture in response to transcriptional perturbation [35, 42]. Disentangling the molecular mechanisms on a genomic scale, requires the use of the complementary "forward approaches" based on bioinformatic methods (see, e.g., [16, 18–20]) or biophysical modeling (e.g., [17, 43]) to predict nucleosome positions from sequence.

Second, since reverse approaches rely on good statistics, our study is presently limited to the study of average patterns, obtained from a large number of different genes. Of course, many genes could have additional features in their free energy landscape at various positions. Again, these features could be directly encoded, by the intrinsic specificity of the DNA-histone interaction [15, 16, 44], or *in trans*, via competition with other specific DNA-binding proteins, biochemical histone modifications [12, 45], or chromatin remodeling [8]. Such additional features do not necessarily affect the average pattern. However, our study firmly establishes the simple physical model of a Tonks gas with "programmable" boundary conditions as an excellent quantitative 'null model' for nucleosome positioning, which can be used as a reference point to identify specific positioning effects as deviations from it. Such a reverse approach on a gene-by-gene basis will likely be very fruitful once data with sufficient statistics and precision becomes available.

## Materials and Methods

### Read density as proxy for nucleosome density.

The data of Mavrich *et al.* [10] is the basis for our analysis. Mavrich *et al.* extracted nucleosomal DNA from yeast cells and sequenced the DNA stretches obtaining a list of reads which they aligned to the *Saccharomyces cerevisiae* genome. Nearly perfect alignments resulted in a list of reads with start and end coordinates on the Watson or Crick strand, which we obtained from the authors. Assuming a nucleosome width of 147 bp we merged the information from both strands and assigned to each read the putative location of the midpoint of the original nucleosomal DNA sequence (see "Supplementary Information" of Ref. [31]). Originally, some reads were aligned to multiple positions on the genome giving them an artificially high weight. Therefore, we counted the number of alignments for each read (number of reads with same read identifier) and weighted the reads by the reciprocal number of their occurrence. For example, if alignment to the yeast genome resulted in 5 hits, each alignment was weighted by a factor 1/5. The frequency of reads vs. location on the genome defines the read density map serving as our proxy for nucleosome density and is denoted by $\Sigma_{reads}$ below. A small region of the read density map is sketched in Fig. 1A and Fig. S2A.

**Genes, ±1 nucleosome positions and alignments.**

Our list of start and end sites of genes is based on the list of transcribed regions and open reading frames as reported in "Supplementary Research Data" of Ref. [10] (file `Supplementary_Table_S2.xls`). We combined the start sites of transcribed regions (class: *pol II*, subclass: *mRNA*) and the end sites of open reading frames (class *pol II*, subclass: *ORF*) with same 'feature ID' to one 'gene' with a total of 4792 genes. See Figs. S3A and B for statistics of length of the genes and distances between TSSs.

We used alignments of the read density map to the positions of nucleosomes surrounding the nucleosome free region (NFR) at the TSS for a quantitative test of statistical positioning. Since the read density map we used for our analysis does not allow direct annotation of individual nucleosomes, we had to employ the list of identified nucleosomes from the "Supplementary Research Data" of Ref. [10] (file `Supplementary_Table_S1.xls`). We used the definition of the +1 nucleosome as the first nucleosome at or downstream from the transcription start site (TSS) while the -1 nucleosome is defined as the first nucleosome upstream from the TSS. The probability distributions of the ±1 nucleosome's distance to the TSS are peaked at some distance from the TSS (Fig. 3A) such that a slightly different definition of the ±1 nucleosome has no significant effect on the results. Next, we aligned the read density map to the position of these nucleosomes and averaged (Fig. 2 and Fig. 4A).

To test the influence of gene starts and ends close to the ±1 nucleosomes of interest, we additionally created alignments using only genes larger than 2000 bp (Fig. S3C) and using those genes without gene start or end sites within 1000 bp upstream from the TSS (Fig. S3D).

**An alternative proxy for nucleosome density.**

The read density map we derived does not include any correction for sequence bias of micrococcal nuclease (MNase). To test for the importance of such a correction, we performed an alternative analysis towards a nucleosome density around the ±1 nucleosomes. To that end, we exploited the list of nucleosomes as identified by Mavrich *et al.*: Based on the reads aligned to the yeast genome, these authors identified individual nucleosomes using a peak detection algorithm after correcting for MNase bias (see "Supplementary Information" of Ref. [31]). The emerging list of nucleosomes also includes the standard deviation (measure of fuzziness) for each nucleosome ("Supplementary Research Data" of Ref. [10], file `Supplementary_Table_S1.xls`). Interpreting the nucleosome's standard deviation as a cell to cell variation (instead of an experimental error), we represented each nucleosome with assigned standard deviation larger than 3 by a Gaussian distribution with standard deviation given by the nucleosome's standard deviation. This results in an alternative proxy $\Sigma_{peaks}$ for the nucleosome density as sketched in Fig. S2B. Both proxies for nucleosome density, the one based on the raw data (reads) and the one based on processed data (individual nucleosomes), significantly differ locally (compare Figs. S2A and S2B). The corresponding alignments to the +1 and -1 nucleosomes, however, are pratically identical (Fig. S2C) having accounted for genome-wide normalization (997655 reads correspond to 52918 nucleosomes). This indicates that MNase bias correction as performed by Mavrich *et al.* is not essential for our analysis.

As a side-remark note that the proxy $\Sigma_{peaks}$ at first sight should represent nucleosome density without any further normalization. However, repeating parts of our fitting analysis (see below) with $\Sigma_{peaks}$ instead of $\Sigma_{reads}$ revealed that a fit to the Tonks gas model is only possible if we allow for a normalization factor significantly different from unity ($\approx 0.8$), suggesting that the proxy $\Sigma_{peaks}$ underestimates the number of nucleosomes. A possible explanation is that up to 20 percent of the nucleosomes were missed by the peak detection filter applied by Mavrich *et al.*. This explanation appears likely, since not all of the yeast nucleosomes are well positioned, i.e., a significant portion of nucleosomes will not lead to a clear peak in the distribution, given the average over many cells that is taken in the experiment.

**Model details and assumptions.**

*Tonks gas.* — Our one-dimensional gas description of statistical nucleosome positioning uses a continuous genome coordinate $x$, whereas in reality nucleosome positions only take on discrete values, in steps of single base pairs (bp). The continuum limit is convenient and justified as long as the average distance between particles, i.e., the linker length, is relatively large compared to the discretization step (the average linker length is typically in the range of 15 - 70 bp, depending on the organism). The statistical physics of a gas of finite-sized particles in a one-dimensional continuous state space has long been worked out in detail [25–27], but for a self-contained presentation we derive the explicit form of the oscillatory pattern using a simple physical argument. To this end we consider the two-particle distribution function $\rho_2(x, x')$ which measures the probability that a particle is found at $x$ and another particle is found at $x'$. Mathematically,

$$\rho_2(x, x') = \left\langle \sum_{i,j \neq i} \delta(x_i - x)\delta(x_j - x') \right\rangle,$$

where $\langle \ldots \rangle$ denotes the average over all possible configurations and $\delta(x)$ denotes the Dirac delta function. In the thermodynamic limit, where the number of particles, $N$, and the length of the interval, $L$, are both large (given an average density $\bar{\rho} = N/L$) and for $x$ and $x'$ far away from the boundaries, $\rho_2(x, x')$ does not depend on $x$ and $x'$ independently, but is only a function of the distance, $\rho_2(x, x') = \rho_2(r)$ with $r = |x - x'|$. To obtain $\rho_2(r)$ explicitly, first note that, with regard to the spaces between particles, a Tonks gas of $N$ particles with width $b$ in an interval of length $L$ is equivalent to a one-dimensional gas of point-like particles in an interval of length $L - Nb$ (for clarity, consider periodic boundary conditions). In the bulk, these point particles are randomly distributed, such that the gap size between neighboring particles has an exponential distribution $p_1(r) = \exp(-r/d)/d$ with mean $d = (L - Nb)/N$. For this gas, the probability $p_k(r)$ of finding the $k$-th neighbor particle at distance $r$ is equivalent to the probability that the sizes of $k$ neighboring gaps sum up to $r$. Since the gap sizes are independent, $p_k(r)$ can be expressed as a convolution of $p_1(r)$ distributions and the Laplace transform $\hat{p}_k(s)$ factorizes, $\hat{p}_k(s) = (\hat{p}_1(s))^k$. Inverse Laplace transformation yields

$$p_k(r) = \frac{r^{k-1} \exp(-r/d)}{d^k \cdot (k-1)!},$$

which is also referred to as the Erlang distribution. The corresponding function for the original Tonks gas is then obtained by reintroducing the particle width $b$. This only amounts to shifting the distance $r$ in $p_k(r)$ by $kb$ and assuring that the resulting function is identical to zero for $r < kb$ by use of the Heaviside step function $\Theta(r)$. The probability of finding the $k$-th particle at a distance $r$ then becomes

$$p_k(r - bk)\,\Theta(r - bk) =: q_k(r)/\bar{\rho},$$

where we introduced the function $q_k(r)$ and used $d \equiv 1/\bar{\rho} - b$. The two-particle distribution function is obtained by multiplying the density of the first particle, $\bar{\rho}$, with the probability to find any particle at distance $r$, regardless of $k$, which amounts to the sum

$$\rho_2(r) = \sum_{k=1}^{\infty} q_k(r).$$

The first few terms $q_k(r)$ are displayed together with the total sum $\rho_2(r)$ in Fig. S1A using typical parameters for nucleosomes. Note that the density $\rho(r)$ of particles close to a boundary particle with perfectly fixed position is simply $\rho_2(r)/\overline{\rho}$. Thus, $q_k(r)/\overline{\rho}$ is the probability of finding the $k$-th particle at a distance $r$ from the boundary. It is interesting to observe that the distance between the maxima of the oscillatory pattern shown in Fig. S1A differs from $1/\overline{\rho}$. This difference is significant only at smaller densities as plotted in Fig. S1B. Note, however, that the average position of the $k$-th particle does not coincide with the maxima of $q_k(r)$ or the maxima of $\rho(r)$, but is simply $k/\overline{\rho}$.

Physically, the oscillations of $\rho_2(r)$ and equally of $\rho(r)$ are a signature of a collective effect, which results from an interplay between the excluded volume interaction and entropy. Very close to a given particle ($r < b$), there is a "depletion layer" which no particle midpoint can access, hence $\rho_2(r) = 0$ for $r < b$. Then, only the leftmost particle can access the first layer $b < r < 2b$. The further this first particle moves to the right, the further it compresses the remainder of the gas. In reaction, the gas exerts a pressure onto the first particle to stay close to the boundary particle, and hence $\rho_2(r) = q_1(r)$ decays within the first layer. However, in the second layer ($2b < r < 3b$), both $q_1(r)$ and $q_2(r)$ contribute and $\rho_2(r)$ increases again. Finally, an oscillatory pattern of $\rho_2(r)$ emerges from summing the individual peaked functions $q_k(r)$ (which are non-zero for $r > kb$ only and decaying for large $r$). The peaks in $\rho_2(r)$ wash out with increasing $r$ since the individual $q_k(r)$ become broader and more $k$ values contribute. The limiting value of $\rho_2(r)$ is $\overline{\rho}^2$, i.e., the square of the mean density. With increasing mean density ($\overline{\rho} \to 1/b$), the individual $q_k(r)$ become sharper and overlap less; oscillations in $\rho_2(r)$ become more pronounced and range further.

*Different boundary conditions.* — Eq. (1) provides an analytic expression for the particle density close to a perfect boundary. This expression can in fact be interpreted and utilized in two different ways: (i) The origin, $r = 0$, can be interpreted as the location of a perfectly positioned nucleosome, which then acts as a perfect boundary for the neighboring nucleosomes. (ii) The origin can be the location of a barrier of another type, e.g., a nucleosome-repelling DNA sequence or bound transcription factors and only the series of peaks for $r > 0$ correspond to nucleosomes. The difference amounts to a horizontal shift: in the former case the $r = 0$ point of the theoretical pattern must be aligned with the first nucleosome, whereas in the latter case the $k = 1$ peak must be aligned with the first nucleosome. This simple shift switches between the two opposite extremes in the range of possible boundary conditions, i.e., perfect direct positioning vs. pure indirect positioning against a perfect barrier. For our quantitative data analysis we limited ourself to these two extreme cases (see 'Procedure for quantitative analysis' below), however in Fig. 4B we also explored the effect of more realistic conditions where neither perfect attraction of a nucleosome to a single point on the genome occurs (e.g., note the finite width of the peak associated with the +1 nucleosome in Fig. 2) nor perfect repulsion. To generate Fig. 4B, we numerically determined the particle density close to a broad repellent region of varying strength, and also close to a narrow attractive region of varying depth and finite width (binding energy here is defined to act on the particle midpoint). We computed the density for a grand-canonical ensemble, using a recursion relation of the same type as described in Ref. [43], and with the chemical potential adjusted such that an average inter-particle spacing of $1/\overline{\rho} \approx 175$ bp was obtained.

*Model assumptions.* — As stated in the main text, our application of the Tonks gas model to the nucleosome data is based on a number of simplifying assumptions. For instance, we assumed that the variation in nucleosome position indicated by the distribution of reads is a true reflection of the cell-to-cell variability. In practice, the nucleosome positions inferred from the reads have some (unknown) experimental error. However, *a posteriori* our assumption appears reasonable, due to the quantitative agreement of model and data, which suggests that the decaying oscillations genuinely reflect the many-body physics of the Tonks gas – such an agreement is not expected if the variation were merely experimental error. Another assumption, shared with basically all models for nucleosome organization, is the equilibrium assumption made by associating the nucleosome distribution with a static free energy landscape. *In vivo*, transcription, DNA replication, and active remodeling processes regularly translocate and evict nucleosomes, and it is questionable to what extent these processes can be captured by a static free energy landscape. Though little is known about the kinetics of chromatin reorganization, we can consider some simple scenarios to illustrate that this assumption may not be as bad as it seems: For instance, remodeling enzymes that merely increase the mobility of nucleosomes, without preference for a certain direction or position, would only speed up the equilibration in a free energy landscape, but not affect its shape. If the remodelers do have any sort of bias, but work rapidly, their effect can be included into a modified free energy landscape. Other passive (competitive binding) and active (repositioning) processes can similarly be included in an effective free energy landscape, as long as their kinetics is rapid on the timescale of interest. Remodelers may also modify the interaction potential between the nucleosomes, beyond the simple hard-core repulsion of the Tonks model. Other effects, including transient unwrapping of the nucleosomal DNA [5, 6, 27] and geometric constraints in higher order structures may modify the interaction between nucleosomes as well. In statistical physics, more complicated interactions between particles in one dimensional gases have been considered [30], however due to the good agreement between the data and the simple Tonks model, we did not consider generalizations in this direction.

**Procedure for quantitative analysis.**

To systematically compare the quantitative model to the $\pm 1$ nucleosome alignments of the read density (i.e., our proxy $\Sigma_{reads}$ for nucleosome density), we performed least squares fits using the function

$$f(r) = \lambda \cdot \rho(r - \Delta r), \qquad (2)$$

where $\lambda$ is a normalization factor, $\Delta r$ tests for a possible horizontal offset in the data, and the function $\rho(r)$ from Eq. (1) contains the parameters $\bar{\rho}$ and $b$. In all our fits, the nucleosome width was kept fixed at $b = 147$ bp. We used the offset parameter $\Delta r$ also to distinguish between the two opposite boundary conditions considered for our fits: As explained above, $\Delta r = 0$, corresponds to the direct positioning scenario where the first nucleosome is a fixed barrier for the neighboring nucleosomes, while a shift by one nucleosome width corresponds to the statistical positioning scenario where the boundary is not a nucleosome, but another repellent feature on the genome. (In the latter case, the different genes should in principle be aligned to the location of the boundary, but since this is not possible, our alignment to the first nucleosome is the best alternative.) For each of our fits, one of these two scenarios is imposed by choice of the starting value for $\Delta r$, since each scenario corresponds to a deep "basin" in the least-squares score function. As can be seen from the Tables in the Supporting Material, each best-fit value for $\Delta r$ either clearly corresponds to the direct positioning scenario, $\Delta r \approx 0$, or to the indirect positioning scenario, $-1/\bar{\rho} < \Delta r < -b$. We performed fits to -1 nucleosome alignment data in the same way as for +1 nucleosome alignment data, except that we mirrored the data at the origin. For the fits, we used the data in a range from 200 to 2000 bp downstream from the +1 nucleosome and upstream from the -1 nucleosome, respectively. Altering the fitting range to 200 - 1200 bp had no significant effect on the results. To ensure best possible parameter estimates, we performed each fit 300 times from a wide range of starting parameters. Best fits are shown in Figs. 2A, S4, and 4C (where a peak at $\Delta r$ has been added where applicable to indicate the directly positioned nucleosome, i.e., for the case $\Delta r \approx 0$). The corresponding parameter estimates are displayed in Table S1 where $\delta$ denotes the squared deviation per data point between data and model.

The parameter estimates from fits to the +1 nucleosome alignment are robust against variations in details of the fitting procedure: (i) Fitting to the average over all genes yields almost the same parameter estimates as a fit to an average where only genes larger than 2000 bp are considered. For the latter, nucleosome density is estimated slightly higher due to the slightly further ranging oscillations (Fig. S3C), but it does not significantly differ from the estimate obtained from the alignment including all genes (see Figs. 2A and S4, Table S1). (ii) Randomly partitioning the set of 4792 genes over which the average is performed into four subsets and repeating the fitting analysis yielded almost identical results, see Table S2. (iii) To account for the effect of the residual cell-to-cell variation in the position of the +1 nucleosome, we also performed a fit using a 'convoluted Tonks model', where Eq. (2) (with $\Delta r = 0$) was convoluted with a probability distribution function corresponding to the experimental nucleosome density in the range of $\pm 30$ bp around zero. The first peak downstream from the +1 nucleosome, corresponding to the +2 nucleosome, is much better characterized by this fit (compare Figs. 2A and B) suggesting that cell-to-cell variation of the +1 nucleosome's position is reflected in cell-to-cell variations of the downstream nucleosomes. Yet, parameter estimates are very similar to those obtained from the fit without convolution (Table S1) indicating that including this effect is not essential when fitting the Tonks gas model to the data in the range of 200 to 2000 bp as we do everywhere else in this study.

For the fit to just the -1 nucleosome alignment, we used the parameter estimates for nucleosome density $\overline{\rho}$ and normalization $\lambda$ obtained from the fit to the +1 nucleosome alignment. Thus, the only remaining fit parameter here was the offset $\Delta r$ (Table S1), which was started at values $\Delta r < -b$. In order to systematically test alternative scenarios (e.g., direct positioning of the -1 nucleosome and indirect positioning of the +1 nucleosome), we performed simultaneous fits to both the +1 and -1 alignment data for each of the four possible boundary conditions. Fits were carried out analogously to the procedure described above, but with the $\lambda$ and $\overline{\rho}$ parameters constrained to take the same values on both sides. Fig. S5 displays the results, Table S3 shows the parameter estimates. Regarding the mean squared deviation per data point $\delta$, scenarios C and D are similar, while scenarios A and B are less probable. In both eligible scenarios, the -1 nucleosome is indirectly positioned. In the best fit scenario C the +1 nucleosome is directly positioned.

**Comparison of average nucleosome densities.**

In our systematic fitting procedure described above, we assumed the same average nucleosome density up- and downstream from the NFR. This must be justified by comparing the average density in intergenic regions to that in genic regions. To estimate their ratio, we used the proxies for nucleosome density described above, i.e., the read density ($\Sigma_{reads}$) and the representation of nucleosomes by Gaussians with appropriate width ($\Sigma_{peaks}$). To exclude the influence of the 5' NFR, which is mostly located within intergenic regions, we excluded the NFR regions. Using proxy $\Sigma_{reads}$ we obtained a ratio of 1.00 for the density in intergenic to the density in genic regions, whereas a ratio of 0.85 resulted from using $\Sigma_{peaks}$. We conclude that there is no clear indication of a density bias between intergenic and genic regions (apart from the existence of NFRs). We therefore assumed equal average density up- and downstream from the TSS for the fitting procedure.

# Acknowledgments

We are grateful to Frank Pugh and Cizhong Jiang for providing the data and many useful comments. We also thank Ho-Ryun Chung and Jonathan Widom for valuable discussions. This work was supported by the German Excellence Initiative via the program Nanosystems Initiative Munich (NIM). W.M. acknowledges funding by the Studienstiftung des deutschen Volkes and the Elite Network of Bavaria via the International Doctorate Program NanoBioTechnology (IDK-NBT). The funders had no role in study design, data collection and analysis, decision to publish, or preparation of the manuscript.

# Figures

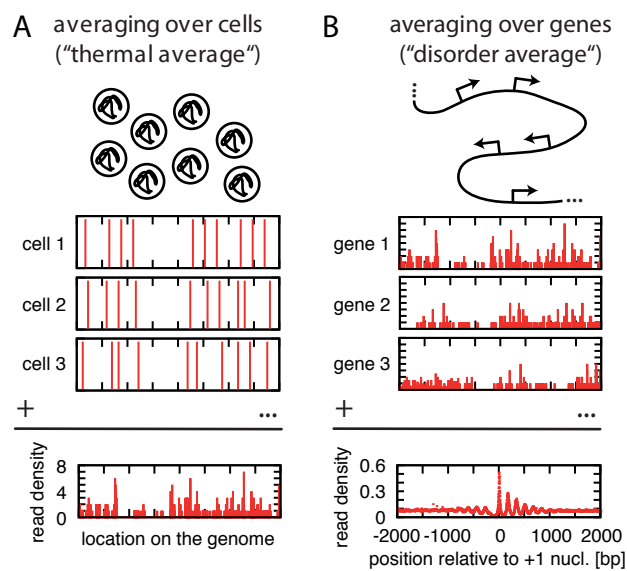

**Figure 1. Illustration of the nature of the available data and its analysis.** (A) Nucleosomal DNA from different cells is extracted and sequenced. The genomic positions of the sequence reads are determined, resulting in a genome-wide density of reads. This map reflects the nucleosome density averaged over an ensemble of cells. Physically, this average is akin to a thermal average. (B) To extract typical features (and to improve the statistics) genes are aligned according to a specific feature (here: the most likely position of the +1 nucleosome), and the read density is averaged over all genes. Physically, this average is akin to a disorder average.

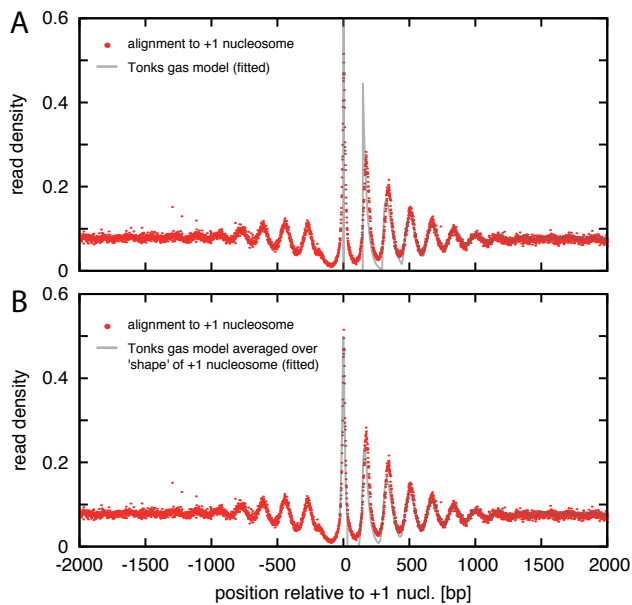

**Figure 2. Comparison of the downstream nucleosome density pattern with the physical model.** (A) The read density (red dots; aligned to the +1 nucleosome locations and averaged over all genes) is displayed together with the best fit by the Tonks model (gray line; least-squares fit between base pairs 200 and 2000, parameters: $b = 147$ bp and $1/\overline{\rho} = 177$ bp). (B) Same as in (A), but with the fit based on a convoluted Tonks gas model which takes into account the finite width of the experimental +1 peak, by convoluting the Tonks gas distribution function with the experimental probability distribution for the +1 peak in the range from $-30$ to $30$ bp (parameters: $b = 147$ bp and $1/\overline{\rho} = 175$ bp).

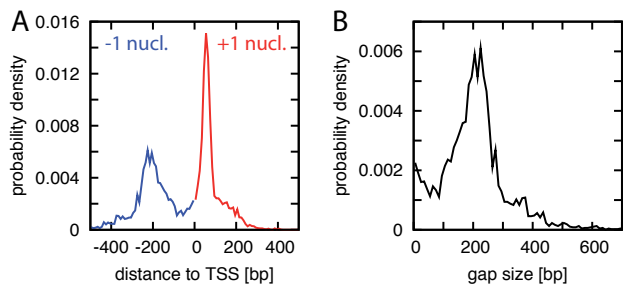

**Figure 3. Distribution of +1 and -1 nucleosome positions.** (A) Probability distribution of the distance of the +1 and -1 nucleosomes to the TSS, obtained as described in 'Materials and Methods'. While the +1 nucleosome is typically found about 50 bp downstream from the TSS, the position of the -1 nucleosome is significantly more disperse. (B) Probability distribution of the gap size, i.e., the distance between the borders of the +1 and -1 nucleosomes given a nucleosome width of 147 bp.

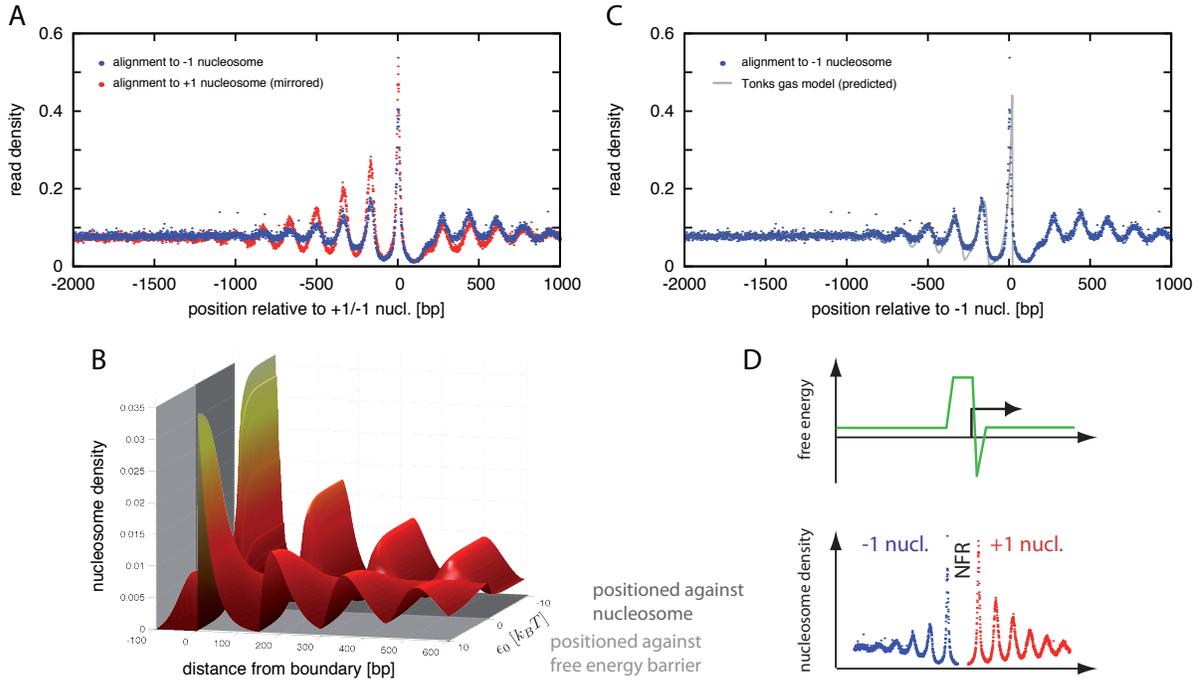

**Figure 4. The upstream pattern and the effect of boundary conditions on statistical positioning.** (A) Comparison of the upstream pattern in the read density (blue dots; all genes aligned to the position of their -1 nucleosome) with the (mirrored) downstream pattern of Fig. 2 (red dots). The patterns are qualitatively similar, but quantitatively significantly different. (B) 3D plot displaying the dependence of the theoretically calculated pattern on the boundary condition. The boundary condition is parameterized by the energy scale $\epsilon_0$ (measured in units of $k_B T$), with $\epsilon_0 > 0$ (light gray shaded region) representing a nucleosome repellent region, while $\epsilon_0 < 0$ (dark gray) describes an attractive potential for a nucleosome (the width of which is chosen here to roughly correspond to the finite peak width of the +1 nucleosome in the data). Parameters are $b = 147$ bp and $1/\bar{\rho} \approx 175$ bp, see 'Materials and Methods' for details. (C) Comparison of the upstream pattern (blue dots) to the Tonks model with boundary condition for a perfectly repellent region with $\epsilon_0 \gg 1$ (gray line; same nucleosome density and normalization as in Fig. 2A). (D) Illustration of the typical nucleosome organization around TSSs and its origin based on the conclusions of the present study. A broad repelling region combined with a localized attractive feature in the free energy landscape close to the TSS (top) leads to a NFR and a directly positioned +1 nucleosome (bottom). The NFR together with the +1 nucleosome form local boundaries which statistically position the nucleosomes in the vicinity, over ranges up to $\sim 1000$ bp.

# Supporting Material for "Quantitative test of the barrier nucleosome model for statistical positioning of nucleosomes up- and downstream of transcription start sites"

Wolfram Möbius and Ulrich Gerland

### Figure S1

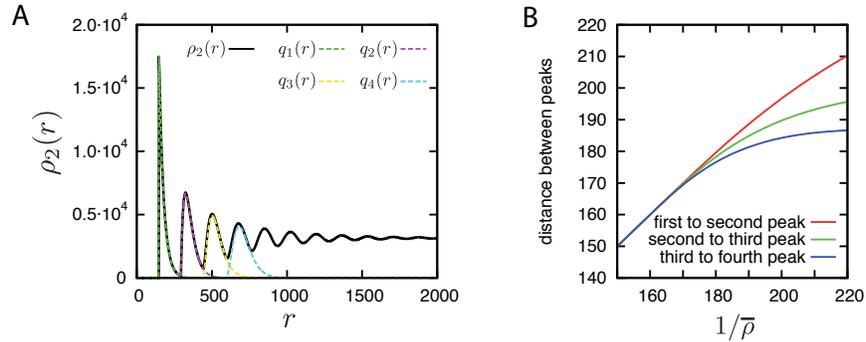

FIG. S1: Characteristics of the Tonks gas two-particle distribution function. (A) Two-particle distribution function $\rho_2(r)$ for a particle size of $b = 147$ and an average particle spacing $1/\bar{\rho} = 178$. The first few individual terms $q_k(r)$ contributing to $\rho_2(r)$ are superimposed. (B) Distance between the individual peaks in $\rho_2(r)$ as a function of $\bar{\rho}$ for $b = 147$. For dense packing, the first few maxima are equidistantly spaced by $1/\bar{\rho}$. Note that the first peak is always located at $r = b$, regardless of the particle density.

### Figure S2

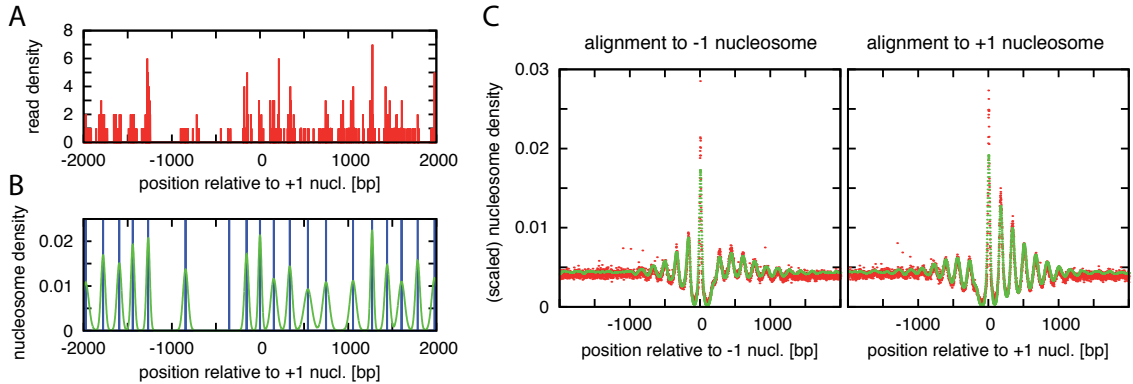

FIG. S2: Comparison between two proxies for the nucleosome density. (A) Section of the read density map ($\Sigma_{reads}$) based on sequence reads reported by Mavrich *et al.* [10] (Genome Res 18:1073–83 (2008)). (B) Section of nucleosome density estimate based on the list of nucleosomes identified by Mavrich *et al.* ($\Sigma_{peaks}$): Each nucleosome is represented by a Gaussian with mean and standard deviation corresponding to the values reported. (C) Alignment of both nucleosome density proxies (red dots for $\Sigma_{reads}$, green dots for $\Sigma_{peaks}$) to $\pm 1$ nucleosome positions and averaging over all genes leads to nearly identical results. To account for the unknown normalization, we scaled the read density map such that the genome-wide number of reads equals the genome-wide number of identified nucleosomes.

**Figure S3**

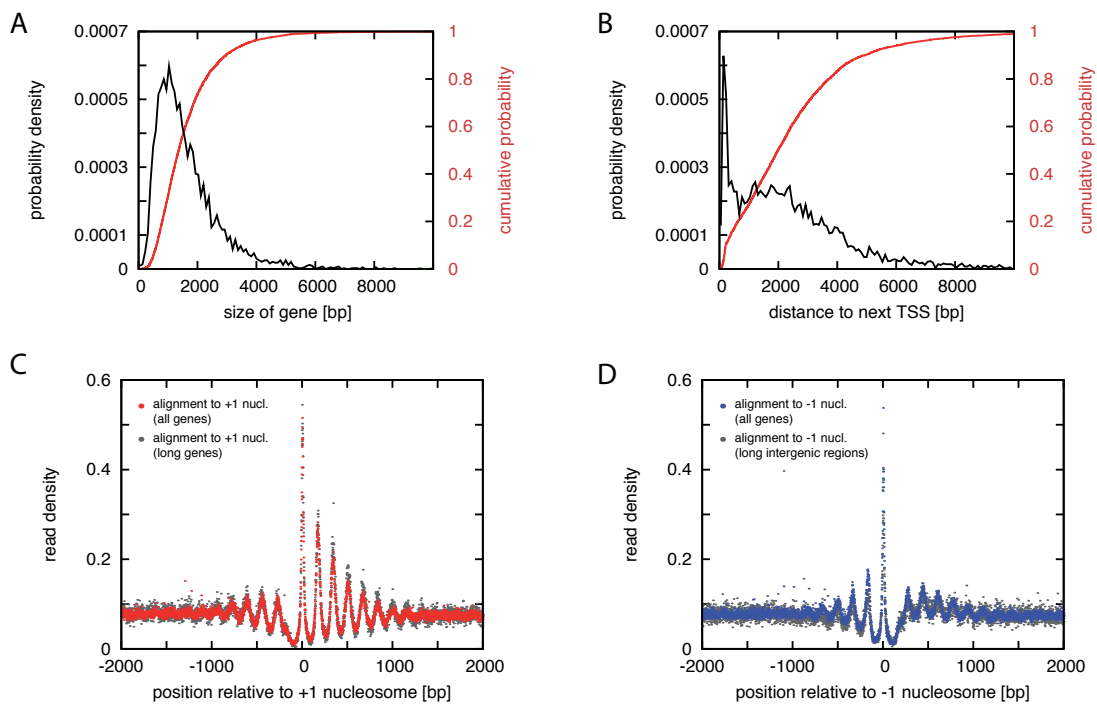

FIG. S3: Distribution of gene start and end sites and effects on alignments. (A) Probability distribution (black) and cumulative distribution (red) for the length of genes. Typical sizes of genes are about 1000 bp, but about nearly a third is larger than 2000 bp. (B) Same distributions for the distance between neighboring TSSs. Distances are in general comparable to the size of genes, but a number of TSSs are very close to each other. (C) Alignment of read density to +1 nucleosome and average over all genes (red dots) and those 1269 genes being larger than 2000 bp only (gray dots). The averages are very similar, but close inspection shows that amplitudes are slightly larger and oscillations range further when considering large genes only. (D) Alignment of read density to -1 nucleosome and average over all genes (blue dots) and those 952 genes where no gene starts or ends were found within 1000 bp upstream of the TSS (gray dots). The averages are very similar, but amplitudes are slightly smaller when considering those genes without other gene starts or ends upstream only.

**Figure S4**

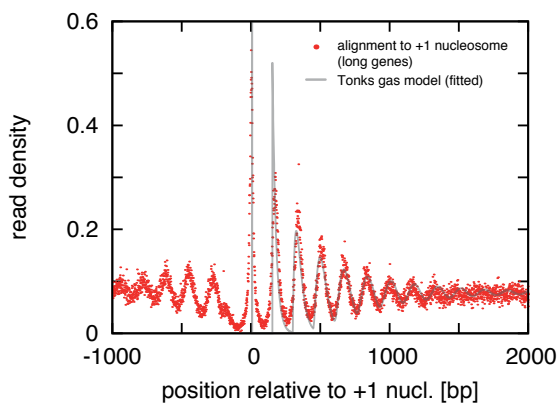

FIG. S4: Best fit of Tonks gas model (gray line) to +1 nucleosome alignment of read density including genes larger than 2000 bp only (red dots). Visual inspection yields good agreement between model and data, comparable to the analogous fit to the data including all genes (Fig. 2A, see also Fig. S3C). For estimated parameters see Table S1.

**Figure S5**

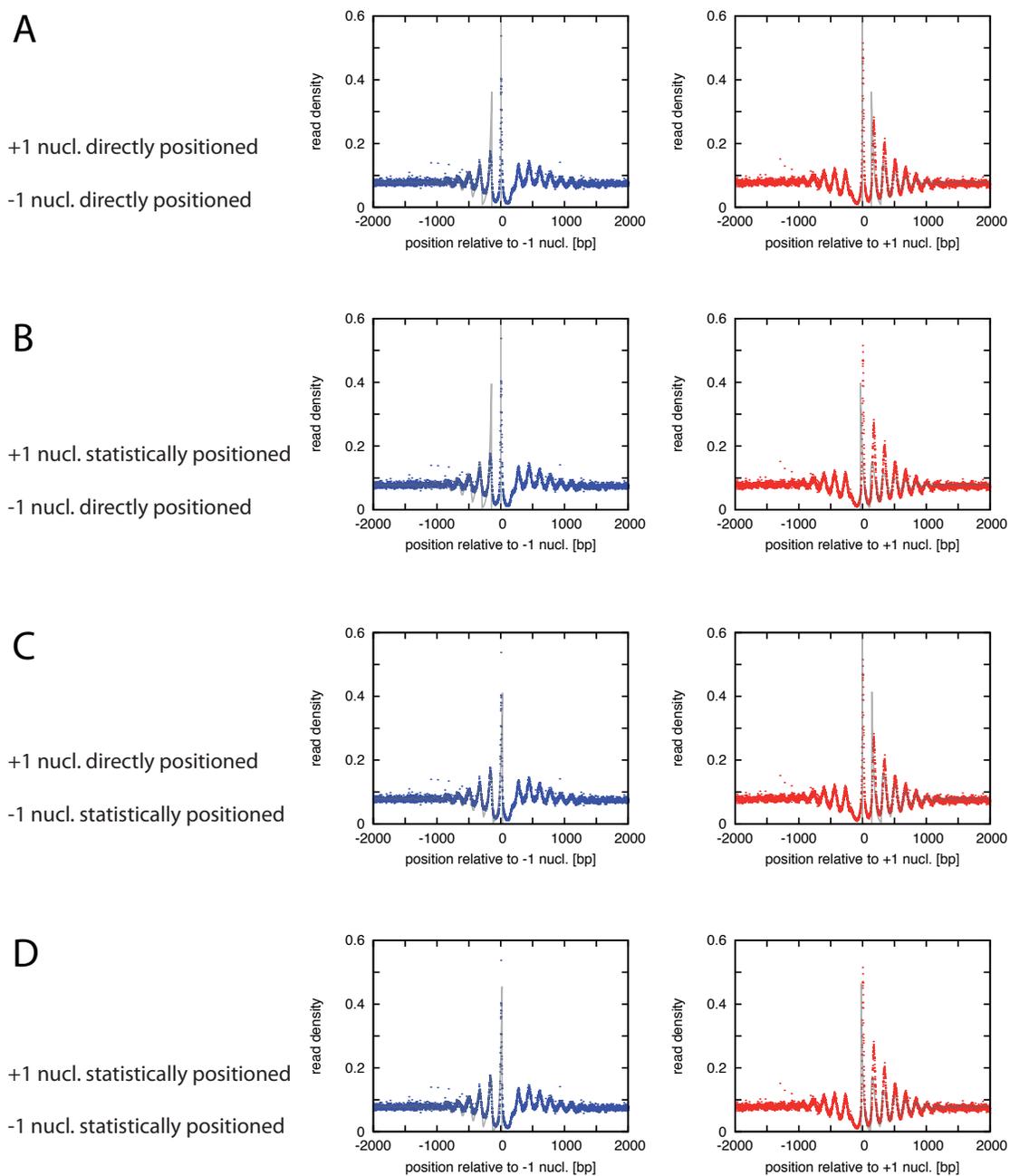

FIG. S5: Best simultaneous fits (gray lines) to -1 and +1 nucleosome alignments of read density (blue and red dots, respectively) given the four possible boundary conditions (both the +1 and -1 nucleosome may be directly or indirectly positioned) with nucleosome density and normalization being equal for both alignments (see 'Materials and Methods' for details). Regarding the mean squared deviation per data point, scenario C describes the data best, i.e., the scenario where the +1 nucleosome is directly positioned while the -1 nucleosome is indirectly positioned (Table S3).

**Figure S6**

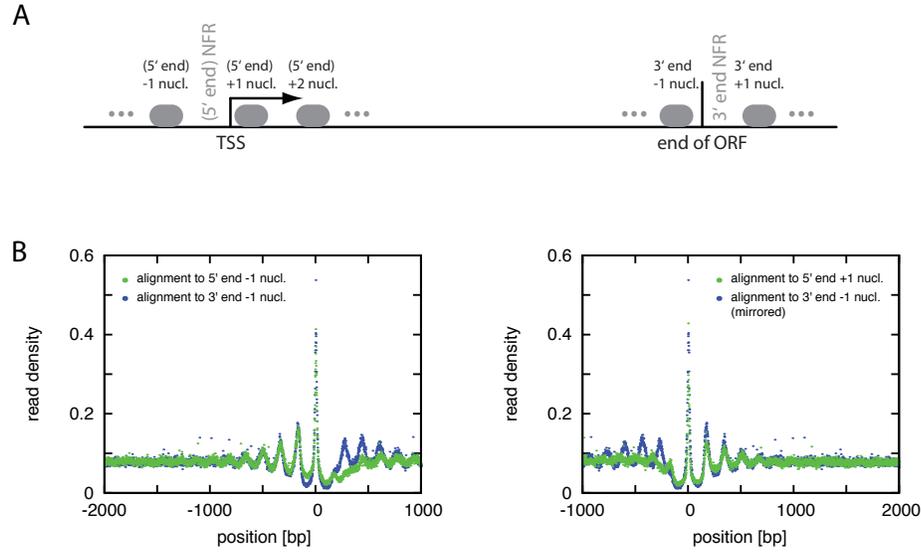

FIG. S6: Nucleosome organization around the 3' end of genes. (A) Sketch of a typical nucleosome organization around both the 5' and 3' ends of genes. Throughout this study, the focus is primarily on the 5' NFR with its flanking -1 and +1 nucleosomes. The nucleosomes flanking the 3' NFR are here referred to as the 3' end -1 nucleosome and the 3' end +1 nucleosome. We determined the positions of 3' end ±1 nucleosomes in analogy to the 5' end ±1 nucleosomes: The 3' end -1 nucleosome is defined as the nucleosome at or first nucleosome upstream of the ORF end while the 3' end +1 nucleosome is the first nucleosome downstream. (B) Alignment of read density to the 3' end -1 nucleosome (left) and 3' end +1 nucleosome (right), respectively (green data points). For comparison, the alignment to the 5' end -1 nucleosome is also shown (blue data points, from Fig. 4, mirrored on the right). Overall, a good agreement is visible between the alignments to the nucleosomes flanking the 3' NFR on both sides and the alignment to the 5' end -1 nucleosome. This indicates that at the 3' end the nucleosomes are only statistically positioned against a repulsive barrier, which we found to be the most likely scenario for the pattern upstream of the 5' NFR. (Note the small bump in the read density within the nucleosome depleted region, just downstream of the 3' end -1 nucleosome and upstream of the 3' end +1 nucleosome; it indicates that the identification of 3' NFRs is not perfect or a certain fraction of genes does not display a 3' NFR.)

## Table S1

| Fit scenario | $1/\overline{\rho}$ [bp] | $\lambda$ | $\Delta r$ [bp] | $\delta$ |
|---|---|---|---|---|
| +1 nucleosome, all genes, Fig. 2A | 177 | 13.7 | 1 | 1.5e-4 |
| +1 nucleosome, genes larger 2000 bp, Fig. S4 | 172 | 13.5 | 7 | 2.8e-4 |
| +1 nucleosome, all genes, fit of convoluted funct., Fig. 2B | 175 | 13.6 | (0) | 1.1e-4 |
| -1 nucleosome, all genes, Fig. 4C | (177) | (13.7) | -169 | 9.8e-5 |

TABLE S1: Parameter estimates from independent fits of Tonks gas model to ±1 nucleosome alignments of read density based on Equation (2) (density $\overline{\rho}$, normalization $\lambda$, offset $\Delta r$, squared deviation per data point $\delta$). Numbers in parentheses indicate values that were set fixed rather than estimated from the fit. See 'Materials and Methods' for details.

## Table S2

| | $1/\overline{\rho}$ [bp] | $\lambda$ | $\Delta r$ [bp] | $\delta$ |
|---|---|---|---|---|
| Partitioning A, Subset 1 | 176.8 | 13.60 | 0.6 | 2.29e-4 |
| Partitioning A, Subset 2 | 177.5 | 13.66 | -0.4 | 2.15e-4 |
| Partitioning A, Subset 3 | 176.9 | 13.65 | 1.3 | 2.46e-4 |
| Partitioning A, Subset 4 | 177.5 | 13.74 | 0.6 | 2.45e-4 |
| Partitioning B, Subset 1 | 176.6 | 13.79 | 1.6 | 2.30e-4 |
| Partitioning B, Subset 2 | 176.6 | 13.65 | 0.6 | 2.54e-4 |
| Partitioning B, Subset 3 | 177.9 | 13.53 | -0.4 | 2.14e-4 |
| Partitioning B, Subset 4 | 177.6 | 13.67 | 0.5 | 2.42e-4 |
| Partitioning C, Subset 1 | 177.2 | 13.74 | -0.3 | 2.36e-4 |
| Partitioning C, Subset 2 | 176.6 | 13.60 | 0.7 | 2.31e-4 |
| Partitioning C, Subset 3 | 177.9 | 13.74 | 0.6 | 2.48e-4 |
| Partitioning C, Subset 4 | 177.3 | 13.59 | 0.6 | 2.24e-4 |
| Partitioning D, Subset 1 | 176.9 | 13.65 | 1.4 | 2.26e-4 |
| Partitioning D, Subset 2 | 177.3 | 13.76 | 0.4 | 2.39e-4 |
| Partitioning D, Subset 3 | 177.2 | 13.61 | -0.4 | 2.28e-4 |
| Partitioning D, Subset 4 | 177.0 | 13.60 | 1.5 | 2.36e-4 |

TABLE S2: Parameter estimates (density $\overline{\rho}$, normalization $\lambda$, offset $\Delta r$, squared deviation per data point $\delta$) from fits of Tonks gas model to +1 nucleosome alignments of read density using subsets of genes only. Four times (partitioning A-D), the set of 4792 genes was divided into four equal-sized subsets (subset 1-4) before fitting. Estimated parameters are very similar; see 'Materials and Methods' for details.

## Table S3

| Fit scenario (see Fig. S5) | $1/\overline{\rho}$ [bp] | $\lambda$ | $\Delta r_{-1}$ [bp] | $\Delta r_{+1}$ [bp] | $\delta$ |
|---|---|---|---|---|---|
| scenario A | 187 | 14.6 | -6 | -8 | 1.6e-4 |
| scenario B | 182 | 14.2 | -2 | -179 | 2.1e-4 |
| scenario C | 180 | 14.1 | -174 | -3 | 1.1e-4 |
| scenario D | 177 | 13.8 | -168 | -170 | 1.3e-4 |

TABLE S3: Parameter estimates for simultaneous fits of Tonks gas model to ±1 nucleosome alignments of read density. Both normalization $\lambda$ and nucleosome density $\overline{\rho}$ are constrained to be equal for both alignments. $\Delta r_{+1}$ and $\Delta r_{-1}$ are independent parameters accounting for different boundary conditions. Regarding the mean squared deviation per data point $\delta$, scenario C describes the data best, i.e., the scenario where the +1 nucleosome is directly positioned while the -1 nucleosome is indirectly positioned (Fig. S5). See 'Materials and Methods' for details.